\documentclass[11pt]{article}

\usepackage{color,graphicx}
\usepackage{young}
\usepackage[vcentermath]{youngtab}
\usepackage{amsmath,amssymb,graphicx}
\usepackage{hyperref}
\definecolor{darkred}{rgb}{0.65,0.15,0}
\hypersetup{pdfborder={0 0 0},colorlinks=true,urlcolor=darkred,citecolor=blue,linkcolor=darkred,linktocpage=true}

\usepackage{cite}
\usepackage{amsmath}
\usepackage{amsfonts}
\usepackage{amssymb}
\usepackage{graphicx}%
\usepackage{amsthm}
\usepackage{mathrsfs}
\usepackage[T1]{fontenc}
\usepackage{enumerate}
\usepackage{color}

\def\4diml{four-dimensional}

\def\-1{^{-1}}

\newcommand{\A}{\mathscr{A}}

\newcommand{\C}{\mathscr{C}}
\newcommand{\M}{\mathscr{M}}
\newcommand{\D}{\mathscr{D}}
\newcommand{\G}{\mathscr{G}}

\newcommand{\tG}{\widetilde{\mathscr{G}}}

\makeatletter

\@addtoreset{equation}{section}
\makeatother

\begin{document}

\thispagestyle{empty}

\vspace{5mm}

\begin{center}
{\LARGE \bf Cosmological string backgrounds from super\\[2mm]   Poisson-Lie T-plurality}

\vspace{15mm}
\normalsize
{\large  \bf Ali Eghbali}

\vspace{2mm}
{\small \em Department of Physics, Faculty of Basic Sciences,\\
Azarbaijan Shahid Madani University, 53714-161, Tabriz, Iran}\\
\vspace{4mm}
\verb"Email: eghbali978@gmail.com"\\
\vspace{7mm}
{\small \it The author dedicates this work to the memory of the people who died due to COVID-19}

\vspace{6mm}

\begin{tabular}{p{12cm}}
{\small
We generalize the formulation of Poisson-Lie (PL) T-plurality proposed by R. von Unge [JHEP 07 (2002) 014] from
Lie groups to Lie supergroups.
By taking a convenient ansatz for  metric of the $\sigma$-model in terms of the left-invariant one-forms of
the isometry Lie supergroups $(C^3 +A)$ and $GL(1|1)$ we construct
cosmological string backgrounds, including $(2+1|2)$-dimensional metric, time-dependent dilaton and vanishing torsion,
in a way that they satisfy the one-loop beta-function equations.
Starting from the decompositions of semi-Abelian Drinfeld superdoubles (DSDs) generated by the $({\C}^3 +{\A})$ and $gl(1|1)$ Lie super bi-algebras
we find the conformal duality/plurality chains of $2+1$-dimensional cosmological string backgrounds coupling with two fermionic fields.
In particular, the new backgrounds obtained by the super PL T-plurality remain conformally invariant at one-loop level.
This work can prompt many new insights into supergravity and obviously has interesting mathematical relations
with double field theory.
}
\end{tabular}
\vspace{-1mm}
\end{center}


\setcounter{page}{1}
\newpage
\tableofcontents

 \vspace{5mm}

\section{Introduction}

PL T-duality was proposed by  Klim\v{c}\'{i}k and \v{S}evera \cite{{Klim1},{Klim2},{Klim3}} in 1995 as an extension of the Abelian T-duality \cite{{Buscher1},{Buscher2}}
to non-Abelian groups of isometries.
Already in their papers  they  considered the possibility of what is now called PL
T-plurality but the first explicit formulas for the background and dilaton shift and
a general discussion of all (possibly) conformal PL T-duality chains in three
dimensions were given by R. von Unge \cite{vonUnge}.
He extended the path-integral formalism for PL T-duality to include the
case of Drinfeld doubles (DDs) which can be decomposed into Lie bi-algebras in more than one way.
The possibility to decompose some
DDs into more than two Manin triples enables us to construct more than two
equivalent $\sigma$-models. This idea was explicitly realized
in \cite{vonUnge} and the classical equivalence of the $\sigma$-models was called the PL T-plurality.
Various aspects of the PL T-plurality along with some examples were discussed in
\cite{{Hlavaty1},{Hlavaty2},{Hlavaty3}} (see also \cite{{Hlavaty4},{Hlavaty5}}).
Furthermore, by using a suitable extension of DD it was obtained formulas for the PL T-plurality transformation
in the presence of spectators for the $\sigma$-model background \cite{Hlavaty6}.
The PL T-plurality has also appeared as an $O(n,n)$ transformation in \cite{Sakatani}.
There, it has been shown that the double field theory  equations of motion are indeed satisfied even in the presence
of spectators; moreover, as a concrete example, the PL T-plurality transformation of $AdS_5 \times S^5$ solution has been studied.
Actually, the development of PL T-duality, based on a Drinfeld double, by describing explicitly its embedding into
double field theory was proposed by F. Hassler in \cite{Faik1} (see also \cite{Faik2}).
Lately, by using the PL T-plurals of cosmologies invariant with
respect to non-simple Bianchi groups, it has been shown that \cite{Hlavaty7} the resulting
plural backgrounds together with dilaton field and a vector field ${\cal J}$
satisfy the generalized supergravity equations.
Of course,  before working out this, the non-Abelian T-duals of Bianchi cosmologies were led to generalized supergravity
solutions once the vector field ${\cal J}$ had been identified with the trace of the structure constants \cite{Eoin}.

In this paper we firstly recall the definitions of Manin supertriple and DSD and briefly
explain the construction of PL T-dual $\sigma$-models on Lie supergroups.
We then generalize the formulation of PL T-plurality proposed by R. von Unge \cite{vonUnge} from Lie groups (DDs) to Lie supergroups (DSDs).
We introduce a $\sigma$-model constructing on $(2+1|2)$-dimensional supermanifold ${\M} \approx O \times  G$
where $G$ is a four-dimensional Lie supergroup of the type $(2|2)$, and $O$ as the orbit of ${ G}$ in $\M$
is a one-dimensional space with the time coordinate. Then,
we take a convenient ansatz for metric of the model in terms of the left-invariant one-forms of Lie supergroup $G$.
Notice that $G$ is considered to be as an isometry supergroup of  the metric.
Accordingly, with the help of the isometry Lie supergroups $(C^3 +A)$ and $GL(1|1)$ \cite{B},
we construct the cosmological string backgrounds including $(2+1|2)$-dimensional metric, time-dependent dilaton and vanishing torsion.
These backgrounds are indeed conformally invariant up to the one-loop order.
Our main results are presented in section 4.
It is shown that the resulting backgrounds are equivalent to
the ones of T-dual $\sigma$-models constructing on semi-Abelian DSDs
$ (({\C}^3 +{\A}) , {\cal I}_{_{(2|2)}})$ and $(gl(1|1) , {\cal I}_{_{(2|2)}})$ where ${\cal I}_{_{(2|2)}}$ is four-dimensional
Abelian Lie supergroup of the type $(2|2)$.
Using the formulation of super PL T-plurality  and starting from the aforementioned
decompositions of DSDs
we find the conformal duality chains of $2+1$-dimensional cosmological string backgrounds coupling with two fermionic fields.
Decompositions of such DSDs were classified in \cite{{ER14},{ERnew}} and \cite{ER6}, respectively,  in terms of Manin supertriples.
It turns out that for each of the Lie superalgebras $({\C}^3 +{\A})$ and $gl(1|1)$
there are several Manin supertriples and the possibility of embedding these Manin supertriples into the corresponding DSDs.
The new backgrounds obtained by super PL T-plurality satisfy the one-loop beta-function equations which are the most
important feature of obtained models.

However, we think that the super PL T-plurality, which will here discuss,
can yield insights to supergravity and obviously has interesting mathematical relations
with double field theory. Because if we have a double field theory solution
with an isometry superalgebra ${\G}$, we may find a series of Manin supertriples,
$({\G} , {\cal I})  \cong  ({\G}' , {\G}'')  \cong  ...,$ and then obtain a chain of double field theory solutions.
In Ref. \cite{Sakatani}, starting from the $AdS_5 \times S^5$ solution as a background with the $(5|1)$ symmetry and also using
series of Manin triples corresponding to a single DD,
it has been obtained the family of non-Abelian T-dual solutions between the $AdS_5 \times S^5$ solution and
type IIA generalized supergravity equations of motion one, as well as the PL T-duality of
the type IIB supergravity solution
to the type IIA supergravity one.
Accordingly, one may generalize the formulation of Ref. \cite{Sakatani} to supermanifolds
and then use the super PL T-plurality formulation in order to find the supergravity solutions.
Finally, the proposed generalization exploiting DSDs can be relevant.

\section{The super PL T-duality/plurality}

We shall consider $\sigma$-models without spectator fields,
i.e., with isomorphic supermanifolds targeting a Lie supergroup.
Let's assume  $G $ to be a Lie supergroup and $\G$ its Lie superalgebra. Suppose now that $G$ acts transitively and freely on $\M$,
then $\sigma$-model having target space in the Lie supergroup $G$,
is given by the  following action\footnote{Notice that the $|a|$ denotes the grading
$a$ such that  $|a|=0$ for the bosonic coordinates and $|a|=1$ for the fermionic
ones. We identify the grading indices by the same indices in the power of $(-1)$, that is, we use $(-1)^a$ instead of  $(-1)^{|a|}$; this notation
has used by Dewitt in \cite{D}. Throughout the paper we use this notation (see appendix A).}
\begin{eqnarray}\label{b.1}
S\;=\;\frac{1}{2}\int_{_{\Sigma}}\!d\sigma^{+}
d\sigma^{-}\;(-1)^{^{|a|}} {L_{_+}}{\hspace{-2mm}^{(l)^a}}\;  {E_{_{ab}}}(g)\;{L_{_-}}^{{\hspace{-2mm}(l)^b}},
\end{eqnarray}
where $\sigma^{\pm} = \tau \pm \sigma$ are the standard light-cone variables on the worldsheet $\Sigma$, and
${L_{\pm}}{\hspace{-2mm}^{(l)^a}}$ are components of the left-invariant Maurer-Cartan one-forms with
left derivative which are defined by means of an element $g: \Sigma \rightarrow G$ in the following formula\footnote{From
now on we will omit the superscript $(l)$ on ${L_{\pm}^{(l)}}^a$.}
\begin{eqnarray}\label{b.2}
g^{-1} \partial_{_{\pm}} g =(-1)^a ~{L_{_{\pm}}^{a}}~T_{{_a}},
\end{eqnarray}
in which $T_{{_a}}, a=1,...,dim \hspace{0.4mm}G$ are the bases of Lie superalgebra ${\G}$, and
${E_{_{ab}}}(g)$  is a certain bilinear form on the ${\G}$, to be specified below.

As noted explicitly in \cite{{Klim1},{Klim2}}, the algebraic structure underlying PL T-duality
is the DD. In the super case \cite{{ER2},{ER5}}, the PL T-dual $\sigma$-models
are also constructed by means of DSDs.
A DSD is simply a Lie supergroup $D$ whose Lie superalgebra $\D =(\G , {\tilde \G})$
admits a decomposition $\D =\G \oplus {\tilde \G}$ into a pair of sub-superalgebras maximally isotropic
with respect to a supersymmetric ad-invariant non-degenerate bilinear form $<.~,~.>$.
Any such decomposition written as an ordered set $(\D  , \G  , {\tilde \G})$
is called Manin supertriple.

The matrix $E(g)$ for $\sigma$-model \eqref{b.1} is of the form\footnote{Here one must use the superinverse formula which has introduced in \cite{D}.}
\begin{eqnarray}\label{b.3}
{ E(g)}\;=\;\big({E_{_0}}^{-1} + \Pi(g)\big)^{-1}, ~~~~~~~~~\Pi(g)=b(g)  a^{-1} (g),
\end{eqnarray}
where ${E_{_0}}$ is a constant matrix,  $\Pi(g)$ defines the super Poisson structure on the Lie supergroup $G$,
and sub-matrices $a(g)$  and $b(g)$ are defined as\footnote{``st'' denotes supertransposition.}
\begin{eqnarray}\label{b.4}
g^{-1} T_{{_a}}~ g &=&(-1)^c ~a_{_{a}}^{^{~c}}(g) ~ T_{{_c}},\nonumber\\
g^{-1} {\tilde T}^{{^a}} g &=&
(-1)^c ~b^{^{ac}}(g)~ T_{{_c}}+{(a^{^{-st}})^{{~a}}}_{c}(g)~{\tilde T}^{{^c}},
\end{eqnarray}
where ${\tilde T}^{{^a}}$ are elements of dual bases in the dual Lie superalgebra ${\tilde \G}$.
The dimension of sub-superalgebras have to be equal
and one can choose a basis in each of the sub-superalgebras
$T_{{_a}} \in \G$  and ${\tilde T}^{{^a}} \in {\tilde \G}$ such that
\begin{eqnarray}\label{b.5}
<T_{{_a}} ,  T_{{_b}}> &=&0,~~~~~~~~~~<{\tilde T}^{{^a}} ,  {\tilde T}^{{^b}}> =0,\nonumber\\
<{\tilde T}^{{^a}}  , T_{{_b}}> &=&{\delta}^{^{a}}_{{~b}}  = (-1)^{ab}~<T_{{_b}} , {\tilde T}^{{^a}}>.
\end{eqnarray}
The generators of the two sub-superalgebras
satisfy the commutation relations
\begin{eqnarray}\label{b.6}
[T_{_a} , T_{_b}] = {f^c}_{ab} ~T_{_c},~~~~~
[{\tilde T}^{^a} , {\tilde T}^{^b}] = {{\tilde f}^{ab}}_{\; \; \: c} ~{\tilde T}^{^c}.
\end{eqnarray}
One can use the super ad-invariance of $<.~,~.>$ to show the remaining commutation relations must be \cite{ER1}
\begin{eqnarray}\label{b.7}
[T_{_a} , {\tilde T}^{^b}] =(-1)^b~ {{{\tilde f}^{bc}}_{\; \; \;a} ~{T}_{_c} + (-1)^a~{f^b}_{ca} ~{\tilde T}^{^c}}.
\end{eqnarray}
It should be noted that the Lie superalgebra structure defined by relations \eqref{b.6} and \eqref{b.7} is called the DSD ${\D}$.
We have used the left-invariant one-forms to write the model in coordinates
${{\cal E}_{_{\mu \nu}}}(x)=(-1)^{^a}~L_{\mu}^{~a} ~{{E}_{_{ab}}}(g)~ {{({{L}^{^{st}}})}^b}_{\nu}$. Thus, we have
\begin{eqnarray}\label{b.8}
S\;=\;\frac{1}{2}\int_{_{\Sigma}}\!d\sigma^{+}
d\sigma^{-}\;(-1)^{^\mu} \partial_+ x^{\mu}\;  {{\cal E}_{_{\mu \nu}}}(x)\;\partial_- x^{\nu},
\end{eqnarray}
where the functions $x^{\mu}: \Sigma \rightarrow \mathbb{R},~\mu =1,...,dim \hspace{0.4mm}G$ are the coordinates of the target supermanifold
which is here isomorphic to the $G$. We say that the background ${{\cal E}_{_{\mu \nu}}}(x)$ has super PL symmetry if \cite{ER2}
\begin{eqnarray}\label{b.9}
{\cal L}_{_{V_{_a}}}{\cal E}_{_{{\mu \nu}}}
=(-1)^{a + \lambda+a \mu+c \rho}~ {{\tilde f}^{bc}}_{\; \; \;a}~{\cal E}_{_{{\mu \rho}}}
{V_{_c}}^{^{\rho}}~{V_{_b}}^{^{\lambda}}~{\cal E}_{_{{\lambda \nu}}},
\end{eqnarray}
for some left-invariant supervector fields $V_a$ satisfying $[V_{_a} , V_{_b}]={f^c}_{ab} ~V_c$.
The integrability condition on the Lie derivative,
$[{\cal L}_{_{V_{_a}}} , {\cal L}_{_{V_{_b}}}]= {\cal L}_{_{[V_{_a} , V_{_b}]}}$,
then implies the mixed super Jacobi identities \cite{ER2} showing that this construction leads naturally to
the DSD.
It is possible to define an equivalent but dual $\sigma$-model by the exchange of $G \leftrightarrow {\tilde G}$,
$\G \leftrightarrow {\tG}$, $E_{_0} \leftrightarrow E_{_0}^{-1}$ and $\Pi(g) \leftrightarrow {\tilde \Pi}({\tilde g}) $.

As noted explicitly in \cite{vonUnge}, the possibility to decompose some
DDs into more than two Manin triples leads to the notion of PL T-plurality. Below,
we shall generalize the formulation of PL T-plurality to the DSDs case and then call it  {\em super PL T-plurality}.
We use the fact that there are, in general, several decompositions (Manin supertriples) of a DSD.
Let $X_{_A}=\{T_{_a} , {\tilde T}^{^b}\},~ a, b=1,...,dim \hspace{0.4mm}G$ be generators of Lie sub-superalgebras $\G$ and $\tG$  of a DSD $\D$
associated with the  $\sigma$-model \eqref{b.1}, and  $X'_{_A}=\{U_{_a} , {\tilde U}^{^b}\}$ are generators
of some other Manin supertriple $({\G}_{_U} , {\tG}_{_U})$ in the same DSD ($\D \cong {\D}_{_U}$) so that
they also satisfy equations \eqref{b.6} and \eqref{b.7}, as well as the bilinear form \eqref{b.5}.
We say that $\D$ and ${\D}_{_U}$ are isomorphic to each other iff there is an invertible supermatrix ${C_{_{A}}}^{^B}$ such that the linear map given by
$X_{_A} = (-1)^{^B}~ {C_{_{A}}}^{^B}~ X'_{_B}$ transforms the Lie multiplication of $\D$ into that of ${\D}_{_U}$ and preserves
the canonical form of the bilinear form $<. , .>$.
Notice that the generators $\{T_{_a} , {\tilde T}^{^b}\}$ define a canonical decomposition of the double.
We define the isomorphism transformation between the $\D$ and  ${\D}_{_U}$ in the matrix form as \cite{vonUnge}
\begin{eqnarray}\label{b.10}
\left( \begin{tabular}{c}
                 T  \\ \hline
                 ${\tilde T}$  \\
                 \end{tabular} \right)=\left( \begin{tabular}{c|c}
                 F & G \\ \hline
                 H & K \\
                 \end{tabular} \right) \left( \begin{tabular}{c}
                 U  \\ \hline
                 ${\tilde U}$  \\
                 \end{tabular} \right).
\end{eqnarray}
One finds the transformation between the bases of $\D$ and  ${\D}_{_U}$ in explicit components form as follows:
\begin{eqnarray}
T_{_a} &=& (-1)^{^c}~ {F_{_a}}^{^c}~U_{_c} + G_{_{ac}}~{\tilde U}^{^c}, \label{b.11}\\
{\tilde T}^{^b} &=& (-1)^{^c}~ H^{^{bc}}~U_{_c} + {K^{^b}}_{_{c}}~{\tilde U}^{^c}.\label{b.12}
\end{eqnarray}
The supergroup associated with the generators $U_{_a}$ will be the supergroup over which the $\sigma$-model is defined. In fact,
the transformed model is then given by the same form as \eqref{b.1} but with
$E(g)$ replaced by \cite{vonUnge}
\begin{eqnarray}\label{b.13}
E(g_{_{_U}})=\big(N M^{^{-1}} + \Pi(g_{_{_U}})\big)^{^{-1}}=\Big((E_{_{0_{U}}})^{-1} + \Pi(g_{_{_U}})\Big)^{^{-1}}.
\end{eqnarray}
One may express $M, N$ and $\Pi(g_{_{_U}})$ in the form of their components as
\begin{eqnarray}
M_{_{ab}}&=& (-1)^{^c}~ {{(K^{st})}_{_{a}}}^{{c}}~ {E_{_0}}_{_{cb}} - {(G^{st})}_{_{ab}},\label{b.14}\\
{N^{^a}}_{_{b}}~&=& {(F^{st})^{a}}_{_{b}}- (-1)^{^c}~ {(H^{st})}^{^{ac}}~ {E_{_0}}_{_{cb}},\label{b.15}\\
\Pi^{^{ab}}(g_{_{_U}})&=& (-1)^{^c}~ b^{^{ac}}(g_{_{_U}}) ~ {{(a^{-1})}_{_c}}^{^b} (g_{_{_U}}),\label{b.16}
\end{eqnarray}
where $a(g_{_{_U}})$ and $b(g_{_{_U}})$ is defined as in \eqref{b.4}
by replacing $g$ with $g_{_{_U}}$ and $T_{_a}({\tilde T}^{^a})$ with $U_{_a}({\tilde U}^{^a})$.
As shown in \cite{vonUnge}, the plurality transformation must be supplemented by a
correction that comes from integrating out the fields on the dual group
in path-integral formulation in such a way that it can be absorbed at the one-loop level into the transformation of the dilaton field.
Following \cite{vonUnge} the formula for the transformation of the dilaton on Lie supergroups is given by
\begin{eqnarray}\label{b.17}
\Phi_{_{_U}} = \phi^{^{(0)}} +\frac{1}{2}\log \big|s\hspace{-0.6mm}\det\big({_{_a}}{E_{_b}}(g_{_{_U}})\big) \big|-
\frac{1}{2}\log \big|s\hspace{-0.6mm}\det\big({_{_a}}{M_{_b}}\big) \big|+
\frac{1}{2}\log \Big|s\hspace{-0.6mm}\det\Big({_{_a}}{(a^{-1})^{^b}} {\tiny (g_{_{_U}})}\Big)\Big|,
\end{eqnarray}
where $\phi^{^{(0)}}$ is the dilaton that makes the original $\sigma$-model conformal and may depend on the coordinates of $G_{_U}$.
The PL T-duality transformation is a special case of the plurality transformation, because one may obtain
the canonical decomposition by choosing
the decomposition matrix \eqref{b.10} as block diagonal, i.e.,
$F=K={\bf 1}, G=H=0$. With this choice one obtains
${N^{^a}}_{_{b}} = {\delta^{^a}}_{_{b}}, M_{_{ab}} = {E_{_{0}}}_{ab}$, then
we arrive at \eqref{b.3}, as well as,
\begin{eqnarray}\label{b.18}
\Phi = \phi^{^{(0)}} +\frac{1}{2}\log \big|s\hspace{-0.6mm}\det\big({_{_a}}{E_{_b}}(g)\big)\big|-
\frac{1}{2}\log \big|s\hspace{-0.6mm}\det\big({_{_a}}{{E_{_{0}}}_{_b}}\big) \big|+
\frac{1}{2}\log \Big|s\hspace{-0.6mm}\det\Big({_{_a}}{(a^{-1})^{^b}} {\tiny (g)}\Big)\Big|.
\end{eqnarray}
From the combination of two formulas  \eqref{b.17} and  \eqref{b.18} one gets the
dilaton transformation between the doubles $\D$ and  ${\D}_{_U}$, giving
\begin{eqnarray}\label{b.19}
\Phi_{_{_U}} &=& \Phi +\frac{1}{2}\log \big|s\hspace{-0.6mm}\det\big({\delta^{^a}}_{_{b}}+\Pi^{^{ac}}(g) {_{_c}}{{E_{_{0}}}_{_b}}\big)\big|-
\frac{1}{2}\log \big|s\hspace{-0.6mm}\det\big({N^{^a}}_{_{b}} +\Pi^{^{ac}}(g_{_{_U}}) {_{_c}}{M_{_b}}\big) \big|\nonumber\\
&&~~~~~~~~-\frac{1}{2}\log \Big|s\hspace{-0.6mm}\det\Big({_{_a}}{(a^{-1})^{^b}} {\tiny (g)}\Big) \Big|
+\frac{1}{2}\log \Big|s\hspace{-0.6mm}\det\Big({_{_a}}{(a^{-1})^{^b}} {\tiny (g_{_{_U}})}\Big)\Big|.
\end{eqnarray}
The super plurality transformation \eqref{b.11}-\eqref{b.16} also includes the canonical dual by choosing the decomposition matrix \eqref{b.10} as block antidiagonal,
that is, for $F=K=0, G=H={\bf 1}$ we get the dual model with $E_{_{0_{U}}} = {\tilde E}_{_0}= {E_{_0}}^{-1}$, corresponding to the interchange
$\G \leftrightarrow \tG$; moreover, one gets the background
${\tilde E}({\tilde g})=\big({E}_{_0} + {\tilde \Pi}({ \tilde g})\big)^{^{-1}}$.

In the following, we will apply the above formulas in order to
construct super PL T-plurals of cosmologies invariant with respect to the $({C}^3 +{A})$ and $GL(1|1)$  Lie supergroups.

\section{Cosmological string backgrounds on supermanifolds}

As in conventional general relativity, homogeneous backgrounds in string cosmology
may be defined as those $3+1$-dimensional spacetime manifolds which admit a $3$-parameter
group of isometries.
It has been shown that \cite{{Ellis},{Callan},{Copeland},{Gasperini}} Bianchi-type string cosmology involves $3+1$-dimensional spatially homogeneous spacetimes
which satisfy at least the lowest-order string beta-function equations.
These string solutions offer prototypes for studying spatial anisotropy and understanding the impacts
of anisotropy on the dynamics of early universe.
Bianchi-type cosmologies can be generally defined in terms of a three-dimensional real Lie
group of isometries that act simply-transitively on three-dimensional, space like orbits
 \cite{{Ellis},{Gasperini}} (see also \cite{{Batakis1},{Batakis2}} and references therein).
These models were then generalized to  $4 + 1$-dimensional cosmological models with four-dimensional real Lie groups whose spatial
hypersurfaces are (simply) connected homogeneous Riemannian
manifolds \cite{Hervik} (see also \cite{Mojaveri}).

Here we present some new solutions of string cosmological models on supermanifolds characterized by $(2+1|2)$-dimensional
metric (with four-dimensional Lie supergroups of the type $(2|2)$ as isometry supergroups), a  dilaton field at most a function of $t$ only and
vanishing torsion.

\subsection{The model setup}

Let us suppose that $\M$ be a (pseudo-)Riemannian target supermanifold of superdimension $(d_{_B} | d_{_F})$
with the coordinates $x^{^M}$, where $d_{_B}$ is the dimension of the bosonic directions,
while $d_{_F}$ denotes the dimension of the fermionic
ones. Notice that because of invertibility of the metric of supermanifold $\M$,  $d_{_F}$ must be even \cite{D}.
Consider the propagation of a bosonic string in the presence of arbitrary backgrounds
of three fields: the supersymmetric metric ${G}_{_{MN}}$, the super antisymmetric tensor field ${B}_{_{MN}}$ and  dilaton $\Phi$
where the labels $M$ and $N$ run from $1$ to $d_{_B} + d_{_F}$.
The string tree level effective action on supermanifold $\M$ for these background fields has the form \cite{ER5}
\begin{eqnarray}\label{c.1}
S_{eff}\;=\;\int d^{^M} x\; \sqrt{G}e^{-2\Phi} \Big[R+4{\overrightarrow{\nabla}_{_M}} {\Phi} \overrightarrow{\nabla}^{^M} {\Phi}
+\frac{1}{12} H_{_{MNP}}H^{^{PNM}}-4\Lambda\Big],
\end{eqnarray}
where $G$ stands for the superdeterminant of $_{_{M}}{G}_{_{N}}$.
The covariant derivative $\overrightarrow{\nabla}_{_M}$ and scalar curvature $R$ are calculated from
the metric ${G}_{_{MN}}$ which is also used for lowering and raising indices and
$H_{_{MNP}}$ defined by\footnote{The relation between the left partial differentiation
and right one has been discussed in appendix A. }
\begin{eqnarray}\label{c.2}
H_{_{MNP}} &=& B_{_{MN}}\frac{\overleftarrow{\partial}}{\partial
x^{^{P}}} +(-1)^{^{M(N+P)}}\; B_{_{NP}} \frac{\overleftarrow{\partial}}{\partial
\Phi^{^{M}}}+(-1)^{^{P(M+N)}}\; B_{_{PM}} \frac{\overleftarrow{\partial}}{\partial
\Phi^{^{N}}}\nonumber\\
&=&(-1)^{^{M}}\;\frac{\overrightarrow{\partial}}{\partial
x^{^{M}}} B_{_{NP}}+(-1)^{^{N+M(N+P)}}\;\frac{\overrightarrow{\partial}}{\partial
x^{^{N}}} B_{_{PM}}+(-1)^{^{P(1+M+N)}}\;\frac{\overrightarrow{\partial}}{\partial
x^{^{P}}} B_{_{MN}},~~~~~~~~
\end{eqnarray}
is the torsion of the field  $B_{_{MN}}$, and $\Lambda$ is a cosmological constant.
The effective action \eqref{c.1} leads to the following equations of motion \cite{ER5}
\begin{eqnarray}
&&{R}_{_{MN}}+\frac{1}{4}H_{_{MQP}} {H^{^{PQ}}}_{_{N}}+2{\overrightarrow{\nabla}_{_M}}
{\overrightarrow{\nabla}_{_N}} \Phi ~=~0,\label{c.3}\\
&&(-1)^{^{P}} {\nabla}^{^P}\big(e^{^{-2 \Phi}} H_{_{PMN}}\big)
~=~0,\label{c.4}\\
&&4 \Lambda-{R}-\frac{1}{12} H_{_{MNP}}H^{^{PNM}} +4{\overrightarrow{\nabla}_{_M}} {\Phi} \overrightarrow{\nabla}^{^M} {\Phi}
 - 4 {\overrightarrow{\nabla}_{_M}} \overrightarrow{\nabla}^{^M} {\Phi}  =0,\label{c.5}
\end{eqnarray}
where ${R}_{_{MN}}$ is the Ricci tensor of the metric (see appendix A). These equations can be also obtained \cite{Callan}
as the conditions of vanishing one-loop beta-functions in the corresponding two-dimensional $\sigma$-model
\begin{eqnarray}\label{c.6}
S &=& \frac{1}{2}\int_{{\Sigma}}\!d\sigma^+  d\sigma^- (-1)^{^M}  ~ \partial_{_+}x{^{M}}\big[G_{_{MN}}(x)
+ B_{_{MN}} (x)\big]
\partial_{_-}x^{{N}} \nonumber\\
&&~~~~~~~~~~~~~~~~~~~~~~~~~~~~~~~~~~~~~~~~~~~~~~~~~~~~- \frac{1}{4 \pi} \int_{{\Sigma}}\!d\sigma^+  d\sigma^-  ~R^{^{(2)}} \Phi(x),
\end{eqnarray}
where $R^{^{(2)}}$ is the worldsheet curvature.

To continue, we consider the action \eqref{c.6} on $(2+1|2)$-dimensional supermanifold ${\M} \approx O \times  G$
where $G$ is a four-dimensional Lie supergroup of the type $(2|2)$ (here we shall use the $(C^3+A)$ and $GL(1|1)$ Lie supergroups
\cite{B}), while $O$ as the orbit of ${ G}$ in $\M$
is a one-dimensional space with time coordinate  $y^{i}=\{t\}$.
We note that the time coordinate does not participate in the PL T-duality transformations
and is therefore called spectator \cite{Sfetsos1}.
As an {\em ansatz} for the metric of the model we consider
\begin{eqnarray}\label{c.7}
ds^2 &= &(-1)^{^{M}}~  dx^{^M}  G_{_{MN}} dx^{^N}\nonumber\\
&& =-dt^2 + (-1)^{^\mu} dx^{\mu}~ g_{_{\mu \nu}}(t,x)  ~dx^{\nu}~~\nonumber\\
&&=-dt^2 + (-1)^{^a} dx^{\mu}~ {{{_{\mu}}{L}}^{a}} ~{E_{_0}}_{ab}(t) ~{({L^{^{st}}})^b}_{\nu}~ dx^{\nu},
\end{eqnarray}
where $x^{\mu}$'s are the coordinates of $(2|2)$-dimensional isometry Lie supergroups with corresponding left-invariant one-forms ${{{_{\mu}}{L}}^{a}}$.
Moreover, we want the tensor field ${B}_{_{MN}}$ to be absent. Then, the action  \eqref{c.6} turns into
\begin{eqnarray}\label{c.6.1}
S &=& \frac{1}{2}\int_{{\Sigma}}\!d\sigma^+  d\sigma^-   ~\Big[-\partial_{_+}t \partial_{_-}t + (-1)^{a} \partial_{_+}x{^{\mu}}
~ {{{_{\mu}}{L}}^{a}} ~{E_{_0}}_{ab}(t) ~{({L^{^{st}}})^b}_{\nu}~ \partial_{_-}x^{{\nu}}\Big] \nonumber\\
&&~~~~~~~~~~~~~~~~~~~~~~~~~~~~~~~~~~~~~~~~~~~~~~~~~~~~- \frac{1}{4 \pi} \int_{{\Sigma}}\!d\sigma^+  d\sigma^-  ~R^{^{(2)}} \Phi(t).
\end{eqnarray}
Notice that  ${E_{_0}}_{ab}(t)$ is, in the presence of the tensor field ${B}_{_{MN}}$, replaced by $E(g,t)$.
For the dilaton field we have considered a cosmological, i.e. time-dependent dilaton.
By a suitable choice of the  matrix ${E_{_0}}_{ab}(t)$ we can construct the metric \eqref{c.7} over a particular Lie supergroup.
We then find a $(2+1|2)$-dimensional metric on the supermanifold ${\M}$ so that ${E_{_0}}_{ab}(t)$ has explicitly not determined yet.
To determine  ${E_{_0}}_{ab}(t)$ and also time-dependent dilaton $\Phi(t)$
we solve the field equations \eqref{c.3}-\eqref{c.5}. In this way, one gets  $(2+1|2)$-dimensional cosmological
string backgrounds which are conformally invariant up to the one-loop order.

\subsection{Cosmological string backgrounds from the $(C^3+A)$ Lie supergroup}

In this subsection we firstly construct the metric \eqref{c.7} using the left-invariant one-forms of the $(C^3 +A)$ Lie supergroup.
Then, we solve the field equations \eqref{c.3}-\eqref{c.5} with the resulting metric
so that the torsion is absent here (i.e., $H_{_{MNP}} = 0$). Before proceeding to construct the metric, let us introduce
the $({\C}^3 +{\A})$ Lie superalgebra \cite{B}.
The $({\C}^3 +{\A})$ is a four-dimensional Lie superalgebra of the type $(2 | 2)$ which has
supersymmetric, ad-invariant and non-degenerate metric on its structure \cite{ER8}.
It is spanned by the set of generators $\{T_{_1}, T_{_2}; T_{_3}, T_{_4}\}$ with gradings;
grade$(T_{_1})$=grade$(T_{_2})=0$ and grade$(T_{_3})$=grade$(T_{_4})=1$, which
fulfill the following  non-zero (anti)commutation rules  \cite{B}:
\begin{eqnarray}\label{c.8}
[T_{_1} , T_{_4}]=T_{_3},~~~~~~~\{T_{_4} , T_{_4}\}=T_{_2}.
\end{eqnarray}
We parametrize an element of the $(C^3+A)$ as
\begin{eqnarray}\label{c.9}
g=e^{\chi T_{_4}} ~ e^{y T_{_1}} ~e^{x T_{_2}} ~e^{\psi T_{_3}},
\end{eqnarray}
where $(y, x)$  are bosonic fields, while $(\psi, \chi)$ are fermionic ones.
Then, using equation \eqref{b.2} we obtain
\begin{eqnarray}\label{c.10}
g^{-1} \partial_{_{\pm}} g = \partial_{_{\pm}} y ~T_{_{1}} + (\partial_{_{\pm}} x-\partial_{_{\pm}} \chi~ \frac{\chi}{2})~T_{_{2}} +
(\partial_{_{\pm}} \psi-\partial_{_{\pm}} \chi~ y)~T_{_{3}}+\partial_{_{\pm}} \chi ~T_{_{4}},
\end{eqnarray}
such that the corresponding left-invariant one-forms components take the following matrix form
\begin{eqnarray}\label{c.11}
_{\mu}{L^{a}} =\left( \begin{array}{cccc}
1  & 0  & 0 & 0\\
0  & 1  & 0 & 0\\
0 & 0  & -1 & 0\\
0  & -\frac{\chi}{2}  & y & -1\\
\end{array} \right).
\end{eqnarray}
In order to make the metric \eqref{c.7} on the $(C^3 +A)$ Lie supergroup one may use the following ansatz
\begin{eqnarray}\label{c.12}
{E_{_0}}_{ab}(t)=\left( \begin{array}{cccc}
a_{_1}(t)  & 0  & 0 & 0\\
0  & a_{_2}(t)  & 0 & 0\\
0 & 0  & 0 & a_{_3}(t)\\
0  & 0  & -a_{_3}(t) & 0\\
\end{array} \right),
\end{eqnarray}
then,
\begin{eqnarray}\label{c.13}
ds^2=-dt^2 +a_{_1}(t) ~dy^2 + a_{_2}(t) ~dx^2 +a_{_2}(t) ~\chi dx d \chi-2  a_{_3}(t) ~ d \psi  d \chi.
\end{eqnarray}
As mentioned above, we shall solve the field equations \eqref{c.3}-\eqref{c.5} for the background including
the metric \eqref{c.13} and a time-dependent dilaton field.
Since the background is torsionless,
the equation \eqref{c.4} is satisfied.
Finally, from equations \eqref{c.3} and \eqref{c.5} the cosmological string background coupled with two fermionic fields is read
\begin{eqnarray}
ds^2 &=& -dt^2 +e^t \Big[dy^2 +dx^2 +\chi ~dx d \chi-2 e^{-2t} d \psi d \chi\Big],\nonumber\\
B&=&0,\nonumber\\
{\Phi} (t) &=& t+\varphi_{_{0}},\label{c.14}
\end{eqnarray}
where $\varphi_{_{0}}$ is an arbitrary constant.
For this solution one finds that the cosmological constant is zero. The metric
is flat in the sense that its scalar curvature is $R=4$.
In addition, one may obtain another solution with a constant dilaton field  and  vanishing cosmological constant in the form
\begin{eqnarray}
ds^2 &=& -dt^2 +{a}_{_0} dy^2 + (t+\frac{b_{_0}}{2})^{^2}~ \big(dx^2 +\chi ~dx d \chi\big) -2 {c}_{_0} d \psi d \chi,\nonumber\\
B&=&0,\nonumber\\
{\Phi} (t) &=& \varphi_{_{0}}.\label{c.15}
\end{eqnarray}
for some constants ${a}_{_0}, b_{_0}$ and $ c_{_0}$.
The metric of this solution is also flat in the sense that its Ricci tensor and scalar curvature vanish.

\subsection{Cosmological string backgrounds from the $GL(1|1)$ Lie supergroup}

Analogously, the $gl(1|1)$ Lie superalgebra  has also supersymmetric, ad-invariant and non-degenerate metric \cite{ER7},
and is defined by the following non-zero (anti)commutation relations
\begin{eqnarray}\label{c.16}
[T_{_1} , T_{_3}]=T_{_3},~~~~~~~[T_{_1} , T_{_4}]=-T_{_4},~~~~~~~\{T_{_3} , T_{_4}\}=T_{_2},
\end{eqnarray}
where $(T_{_1}, T_{_2})$ and $(T_{_3}, T_{_4})$ are bosonic and fermionic bases, respectively.
In Backhouse's classification \cite{B}, the $gl(1|1)$ has been labeled by $({\C}_{-1}^2 + \A)$.
The parametrization of a general element of $GL(1|1)$ we choose as in \eqref{c.9}, giving
\begin{eqnarray}\label{c.17}
g^{-1} \partial_{_{\pm}} g = \partial_{_{\pm}} y ~T_{_{1}} + (\partial_{_{\pm}} x-\partial_{_{\pm}} \chi~ \psi e^y)~T_{_{2}} +
(\partial_{_{\pm}} y~ \psi+\partial_{_{\pm}} \psi)~T_{_{3}}+\partial_{_{\pm}} \chi ~ e^y~T_{_{4}}.
\end{eqnarray}
In this case one may choose the following ansatz
\begin{eqnarray}\label{c.18}
{E_{_0}}_{ab}(t)=\left( \begin{array}{cccc}
a_{_1}(t)  & b(t)  & 0 & 0\\
b(t)  & a_{_2}(t)  & 0 & 0\\
0 & 0  & 0 & a_{_3}(t)\\
0  & 0  & -a_{_3}(t) & 0\\
\end{array} \right).
\end{eqnarray}
Using \eqref{c.17} and \eqref{c.18} together with \eqref{c.7} we obtain
\begin{eqnarray}\label{c.19}
ds^2 &=& -dt^2 +a_{_1}(t) dy^2 + a_{_2}(t) dx^2 + 2 b(t) dy dx +  2\psi e^y \big(b(t)-a_{_3}(t)\big) dy d \chi\nonumber\\
&&~ ~+ 2\psi e^y a_{_2}(t) dx d \chi -2 e^y a_{_3}(t) d\psi d\chi.
\end{eqnarray}
The resulting solution to one-loop beta-function equations is
\begin{eqnarray}
ds^2 &=& -dt^2 - \frac{1}{2} t^2  dy^2 + 2 b_{_0} dy dx -2 b_{_0} e^{y} d \psi d \chi,\nonumber\\
B&=&0,\nonumber\\
{\Phi} (t) &=& {\varphi}_{_0},\label{c.21}
\end{eqnarray}
for some constants $b_{_0}$ and $ {\varphi}_{_0}$. The metric is flat in the sense that its Ricci tensor and scalar curvature vanish.
Moreover, for this solution  we get $\Lambda =0$.
Another solution for which the dilaton field is time-dependent exists and it is given by
\begin{eqnarray}
ds^2 &=& -dt^2 +(t+e^t) dy^2 + 2 b_{_0} dy dx-2 b_{_0} e^{y} d \psi d \chi,\nonumber\\
B&=&0,\nonumber\\
{\Phi} (t) &=& \frac{1}{2} t + {\varphi}_{_0}.\label{c.22}
\end{eqnarray}
For the metric of this solution  one finds that the only non-zero component of the
Ricci tensor is ${R}_{_{yy}}=(1+e^t)/2$, then, ${ R}=0$. Moreover,
the corresponding cosmological constant to this solution is $\Lambda =1/4$.
The resulting conformal backgrounds will be useful in the next section.
We will show that these backgrounds are equivalent to the ones of T-dual $\sigma$-models constructing on semi-Abelian DSDs
$ (({\C}^3 +{\A}) , {\cal I}_{_{(2|2)}})$ and $(gl(1|1) , {\cal I}_{_{(2|2)}})$.

\section{Super PL T-plurality of cosmologies invariant with respect to the $({C}^3 +{A})$ and $GL(1|1)$  Lie supergroups}

In this section, we present forms of the conformally invariant backgrounds with vanishing torsion
and a dilaton field at most a function of $t$ only on semi-Abelian DSDs $\big(({\C}^3 +{\A}) , {\cal I}_{_{(2|2)}}\big)$ and
$\big(gl(1|1) , {\cal I}_{_{(2|2)}}\big)$ in a way that the resulting backgrounds are in agreement with  those of section 3.
Then, using the formulation of super PL T-plurality  we obtain the conformal duality chains of cosmological string backgrounds
with non-vanishing torsion on the other Manin supertriples in the
corresponding DSDs so that they satisfy one-loop beta-function equations.

One can find the classification of real $(4|4)$-dimensional DSDs generated by the $gl(1|1)$
Lie super bi-algebras and their decompositions into six classes of
non-isomorphic Manin supertriples in Ref. \cite{ER6}.
Here we shall present only those occurring in
this paper, i.e., only first class including the following two isomorphic Manin supertriples:
\begin{eqnarray}
\big(gl(1|1) , {\cal I}_{_{(2|2)}}\big) \cong  \big(gl(1|1) , {\C}_{_{p=-1}}^{^{2}} \oplus {\A}_{_{1,1}}.ii\big).\label{d.1}
\end{eqnarray}
In addition, in Ref. \cite{ER14} we have performed complete classification of Lie super bi-algebra structures
on the $({\C}^3 +{\A})$ Lie superalgebra. Lately, we have listed the corresponding $(4|4)$-dimensional
DSDs  including all possible decompositions into the Manin supertriples and
have obtained 24 non-isomorphic classes \cite{ERnew}. Here is one of those classes that we need in
studying the conformal duality chain
\begin{eqnarray}
\big(({\C}^3 +{\A}) , {\cal I}_{_{(2|2)}}\big) \cong  \big(({\C}^3 +{\A}) , {\C}^3 \oplus {\A}_{_{1,1}}.i\big) \cong
\big(({\C}^3 +{\A}) , ({\C}^3 +{\A})_{_{k=4}}^{^{\epsilon=-1}}\big).\label{d.2}
\end{eqnarray}
It should be noted that the label of each Manin supertriple, e.g. $\big(gl(1|1) , {\C}_{_{p=-1}}^{^{2}} \oplus {\A}_{_{1,1}}.ii\big)$,
indicates the structure of the first sub-superalgebra $\G$, e.g.  $gl(1|1)$, the
structure of the second sub-superalgebra $\tG$, e.g.  ${\C}_{_{p=-1}}^{^{2}} \oplus {\A}_{_{1,1}}.ii$; roman numbers $i$, $ii$ etc. (if
present) distinguish between several possible pairings $<. , .>$ of the sub-superalgebras $\G$ and $\tG$,
and the parameter $k$ (e.g. in \eqref{d.2}) indicates the Manin supertriples differing in the rescaling of  $<. , .>$.


\subsection{Conformal duality chain starting from $\big(({\C}^3 +{\A}) , {\cal I}_{_{(2|2)}}\big)$}

\smallskip
$\bullet$~$\big(({\C}^3 +{\A}) , {\cal I}_{_{(2|2)}}\big)$:
\smallskip

\smallskip
In order to obtain a possible conformal duality chain on the non-isomorphic DSDs \eqref{d.2}
we take $\big(({\C}^3 +{\A}) , {\cal I}_{_{(2|2)}}\big)$ as the canonical decomposition
whose Lie superalgebra is given by \cite{ERnew}
\begin{eqnarray}
[T_{_1} , T_{_4}] &=&T_{_3},~~~~~~~~~~~\{T_{_4} , T_{_4}\}=T_{_2},~~~~~~~~~~[T_{_1} , {\tilde T}^{^3}]=-{\tilde T}^{^4},\nonumber\\
{[T_{_4} , {\tilde T}^{^2}]}&=& -{\tilde T}^{^4},~~~~~~~~\{T_{_4} , {\tilde T}^{^3}\}=-{\tilde T}^{^1}. \label{d.3}
\end{eqnarray}
Here and henceforth $(T_{_1}, T_{_2}, {\tilde T}^{^1}, {\tilde T}^{^2})$ and
$(T_{_3}, T_{_4}, {\tilde T}^{^3}, {\tilde T}^{^4})$ are bosonic and fermionic bases, respectively.
As mentioned in section 2, to obtain the canonical decomposition we choose
the decomposition matrix \eqref{b.10} as block diagonal
($F=K={\bf 1}, G=H=0$), then,  we arrive at ${N^{^a}}_{_{b}} = {\delta^{^a}}_{_{b}}$ and $M_{_{ab}} = {E_{_{0}}}_{ab}$.
We use the parametrization of a general group element of $({C}^3 +{A})$ as in \eqref{c.9}, then, one utilizes
\eqref{b.4} in order to calculate the matrices $a(g)$ and $b(g)$ for this decomposition, giving
\begin{eqnarray}\label{d.4}
a_{_{a}}^{^{^{~b}}}(g) =\left( \begin{array}{cccc}
1  & 0  & -\chi & 0\\
0  & 1  & 0 & 0\\
0 & 0  & -1 & 0\\
0  & -{\chi}  & y & -1\\
\end{array} \right),~~~~~~~~~~~~b^{^{ab}}(g)=0.
\end{eqnarray}
Since $b(g)=0$ there is no super antisymmetric tensor field $B_{_{MN}}$. This enables us to obtain the background \eqref{c.14} from
a $\sigma$-model constructing on the double $\big(({\C}^3 +{\A}) , {\cal I}_{_{(2|2)}}\big)$.
To this end, we have to choose  the matrix $E_{_0}$ in the following form
\begin{eqnarray}\label{d.5}
{E_{_0}}_{ab}(t) =\left( \begin{array}{cccc}
e^t  & 0  & 0 & 0\\
0  & e^t  & 0 & 0\\
0 & 0  & 0 & e^{-t}\\
0  & 0  & -e^{-t} & 0\\
\end{array} \right).
\end{eqnarray}
By making use of \eqref{d.4} it then follows that $\Pi(g) =0$, which gives us $E =E_{_0}$ as expected for the canonical decomposition.
The dilaton field that makes the original $\sigma$-model conformal
is found by using equation \eqref{b.18} to be
$\Phi = \phi^{^{(0)}}$. However, according to \eqref{c.14} since we want the total dilaton to be ${\Phi} (t) = t+\varphi_{_{0}}$ we need to
choose $\phi^{^{(0)}} = t+\varphi_{_{0}}$ in which $\varphi_{_{0}}$ is an arbitrary constant.
On the other hand, the corresponding left-invariant one-forms have been given in equation \eqref{c.11}.
Finally, using \eqref{c.6.1} one can construct the original $\sigma$-model on the double $(({\C}^3 +{\A}) , {\cal I}_{_{(2|2)}})$
whose background is
\begin{eqnarray}
ds^2 &=& -dt^2 +e^t \big[dy^2 +dx^2 +\chi ~dx d \chi-2 e^{-2t} d \psi d \chi\big],\nonumber\\
B&=&0,\nonumber\\
{\Phi} (t) &=& t+\varphi_{_{0}}.\label{d.6}
\end{eqnarray}
Indeed, this background is nothing but \eqref{c.14}.
\\
\\
$\bullet$~$\big(({\C}^3 +{\A}) , {\C}^3 \oplus {\A}_{_{1,1}}.i\big)$:~~~~~~~~~\\
\smallskip
Considering the linear map $X_{_A} = (-1)^{B}~ {C_{_{A}}}^{^B}~ X'_{_B}$ transforming the Lie multiplication of $\D$ into that of ${\D}_{_{U}}$
and preserving the canonical form of the bilinear form $<. , .>$,
we find the following non-trivial decomposition matrix between the doubles $\big(({\C}^3 +{\A}) , {\cal I}_{_{(2|2)}}\big)$
and $\big(({\C}^3 +{\A}) , {\C}^3 \oplus {\A}_{_{1,1}}.i\big)$ \cite{ERnew}
\begin{eqnarray}\label{d.7}
\left( \begin{tabular}{c}
                 $T_{_1}$  \\
                 $T_{_2}$  \\
                 $T_{_3}$  \\
                 $T_{_4}$  \\
                 ${\tilde T}^{^1}$  \\
                 ${\tilde T}^{^2}$  \\
                 ${\tilde T}^{^3}$  \\
                 ${\tilde T}^{^4}$  \\
                 \end{tabular} \right)=\left( \begin{tabular}{cccccccc}
                 1 & 0  & 0 & 0 & 0 & 0 & 0 & 0\\
                 0 & 1  & 0 & 0 & 0 & 0 & 0 & 0\\
                 0 & 0  & 1 & 0 & 0 & 0 & 0 & 0\\
                 0 & 0  & 0 & 1 & 0 & 0 & 0 & 0\\
                 0 & -1  & 0 & 0 & 1 & 0 & 0 & 0\\
                 1 & 0  & 0 & 0 & 0 & 1 & 0 & 0\\
                 0 & 0  & 0 & 0 & 0 & 0 & 1 & 0\\
                 0 & 0  & 0 & 0 & 0 & 0 & 0 & 1\\
                 \end{tabular} \right) \left( \begin{tabular}{c}
                 $U_{_1}$  \\
                 $U_{_2}$  \\
                 $U_{_3}$  \\
                 $U_{_4}$  \\
                 ${\tilde U}^{^1}$  \\
                 ${\tilde U}^{^2}$  \\
                 ${\tilde U}^{^3}$  \\
                 ${\tilde U}^{^4}$  \\
                 \end{tabular} \right),
\end{eqnarray}
where $\{U_{_a} , {\tilde U}^{^b}\}, a, b=1,...,4$ are elements of bases of the double
$\big(({\C}^3 +{\A}) , {\C}^3 \oplus {\A}_{_{1,1}}.i\big)$ whose Lie superalgebra  is defined by
the following non-zero (anti)commutation relations \cite{ERnew}:
\begin{eqnarray}
[U_{_1} , U_{_4}] &=&U_{_3},~~~~~~~~~~\{U_{_4} , U_{_4}\}=U_{_2},~~~~~~~~~~[{\tilde U}^{^2} , {\tilde U}^{^3}]={\tilde U}^{^4},\nonumber\\
{[U_{_1} , {\tilde U}^{^3}]}&=& -{\tilde U}^{^4},~~~~~~~[U_{_4} , {\tilde U}^{^2}]=U_{_3}-{\tilde U}^{^4},
~~~~\{U_{_4} , {\tilde U}^{^3}\}=U_{_2}-{\tilde U}^{^1}. \label{d.8}
\end{eqnarray}
Here and henceforth $(U_{_1}, U_{_2}, {\tilde U}^{^1}, {\tilde U}^{^2})$ are bosonic  bases, while
$(U_{_3}, U_{_4}, {\tilde U}^{^3}, {\tilde U}^{^4})$ are fermionic ones.
One utilizes equations \eqref{b.14} and \eqref{b.15} together with \eqref{d.5} to find the matrices
$M_{_{ab}}$ and ${N^{^a}}_{_{b}}$, obtaining
\begin{eqnarray}\label{d.9}
M_{_{ab}} =\left( \begin{array}{cccc}
e^t  & 0  & 0 & 0\\
0  & e^t  & 0 & 0\\
0 & 0  & 0 & e^{-t}\\
0  & 0  & -e^{-t} & 0\\
\end{array} \right),~~~~~~~~~{N^{^a}}_{_{b}}=\left( \begin{array}{cccc}
1  & -e^t  & 0 & 0\\
e^t  & 1  & 0 & 0\\
0 & 0  & 1 & 0\\
0  & 0  & 0 & 1\\
\end{array} \right).
\end{eqnarray}
Since the first sub-superalgebra is $({\C}^3 +{\A})$, the
left-invariant one-forms and the matrix $a_{_{a}}^{^{^{~b}}}(g_{_U})$ are the same forms as in \eqref{c.11} and \eqref{d.4}, respectively.
Using \eqref{b.4}, \eqref{b.16}, \eqref{c.9} and \eqref{d.8}  we then find that
\begin{eqnarray}\label{d.10}
{\Pi}^{^{ab}}(g_{_U}) =\left( \begin{array}{cccc}
0  & 0  & 0 & 0\\
0  &0  & \chi & 0\\
0 & -\chi  & 0 & 0\\
0  & 0  & 0& 0\\
\end{array} \right).
\end{eqnarray}
These results give us the background
\begin{eqnarray}
{E}_{_{ab}}(g_{_U}) &=& \left( \begin{array}{cccc}
\frac{e^{^{-t }}}{\Delta(t)} & \frac{1}{\Delta(t)}  & 0 & \frac{\chi e^{^{-t }}}{\Delta(t)}\\\vspace{1mm}
-\frac{1}{\Delta(t)}  & \frac{e^{^{-t }}}{\Delta(t)}  & 0 & \frac{\chi e^{^{-2t }}}{\Delta(t)}\\\vspace{1mm}
0 & 0  & 0 & e^{^{-t }}\\\vspace{1mm}
\frac{\chi e^{^{-t }}}{\Delta(t)}  & -\frac{\chi e^{^{-2t }}}{\Delta(t)}  & -e^{^{-t }} & 0\\\vspace{1mm}
\end{array} \right),\nonumber\\
\Phi_{_U} &=&\phi^{^{(0)}} -t -\frac{1}{2} \log\big(\Delta(t)\big),\label{d.11}
\end{eqnarray}
where $\Delta(t) = e^{-2t} +1$.
Finally we get the dilaton by remembering that $\phi^{^{(0)}}= t+\varphi_{_{0}}$ which gives the final result
\begin{eqnarray}\label{d.12}
\Phi_{_U} = \varphi_{_{0}} -\frac{1}{2} \log\big(\Delta(t)\big).
\end{eqnarray}
The supersymmetric part of the matrix ${E}_{_{ab}}(g_{_U})$ gives the metric in the coordinate basis,
whereas the super antisymmetric part of ${E}_{_{ab}}(g_{_U})$ gives the super antisymmetric tensor.
Thus, the background in the coordinate basis is read off
\begin{eqnarray}
ds^2 &=& -dt^2 +\frac{e^{^{-t }}}{\Delta(t)} \Big[dy^2 +dx^2 -2 \chi ~dy d \chi +  \chi ~dx d \chi\Big] -2 e^{-t} d \psi d \chi,\nonumber\\
B&=&\frac{1}{\Delta(t)}\Big[dy \wedge dx +\frac{1}{2} \chi dy \wedge d \chi - \chi e^{-2t} dx \wedge d \chi\Big],\nonumber\\
{\Phi} (t) &=& \varphi_{_{0}} -\frac{1}{2} \log\big(\Delta(t)\big).\label{d.13}
\end{eqnarray}
One immediately finds that the scalar curvature of the metric is $R=6 e^{^{-2t }}/\Delta^2(t)$.
Looking at the one-loop beta-function equations
one verifies the conformal invariance conditions of the  background \eqref{d.13} with vanishing cosmological constant.\\
\\
$\bullet$~$\big(({\C}^3 +{\A}) , ({\C}^3 +{\A})_{_{k=4}}^{^{\epsilon=-1}}\big)$:~~~~~~~~~\\\\
The Lie superalgebra of the double $\big(({\C}^3 +{\A}) , ({\C}^3 +{\A})_{_{k=4}}^{^{\epsilon=-1}}\big)$
obeys the following set of non-trivial (anti)commutation relations \cite{ERnew}:
\begin{eqnarray}
[U_{_1} , U_{_4}] &=&U_{_3},~~~~~~~~~\{U_{_4} , U_{_4}\}=U_{_2},~~~~~~~~~~~~~~[{\tilde U}^{^2} , {\tilde U}^{^3}]=-{\tilde U}^{^4},\nonumber\\
{\{{\tilde U}^{^3} , {\tilde U}^{^3}\}}&=& 4 {\tilde U}^{^1},~~~~~~~~[U_{_4} , {\tilde U}^{^2}]=-U_{_3}-{\tilde U}^{^4},
~~~~\{U_{_4} , {\tilde U}^{^3}\}=-U_{_2}-{\tilde U}^{^1},\nonumber\\
{[U_{_1} , {\tilde U}^{^3}]}&=& -4 U_{_3}-{\tilde U}^{^4}.\label{d.14}
\end{eqnarray}
The isomorphism transformation between the doubles $\big(({\C}^3 +{\A}) , {\cal I}_{_{(2|2)}}\big)$
and $\big(({\C}^3 +{\A}) , ({\C}^3 +{\A})_{_{k=4}}^{^{\epsilon=-1}}\big)$ is given by the following decomposition matrix \cite{ERnew}:
\begin{eqnarray}\label{d.15}
\left( \begin{tabular}{c}
                 $T_{_1}$  \\
                 $T_{_2}$  \\
                 $T_{_3}$  \\
                 $T_{_4}$  \\
                 ${\tilde T}^{^1}$  \\
                 ${\tilde T}^{^2}$  \\
                 ${\tilde T}^{^3}$  \\
                 ${\tilde T}^{^4}$  \\
                 \end{tabular} \right)=\left( \begin{tabular}{cccccccc}
                 1 & 0  & 0 & 0 & 0 & 0 & 0 & 0\\
                 0 & 1  & 0 & 0 & 0 & 0 & 0 & 0\\
                 0 & 0  & 1 & 0 & 0 & 0 & 0 & 0\\
                 0 & 0  & 0 & 1 & 0 & 0 & 0 & 0\\
                 0 & -1  & 0 & 0 & 1 & 0 & 0 & 0\\
                 1 & 0  & 0 & 0 & 0 & 1 & 0 & 0\\
                 0 & 0  & 0 & 2 & 0 & 0 & 1 & 0\\
                 0 & 0  & 2 & 0 & 0 & 0 & 0 & 1\\
                 \end{tabular} \right) \left( \begin{tabular}{c}
                 $U_{_1}$  \\
                 $U_{_2}$  \\
                 $U_{_3}$  \\
                 $U_{_4}$  \\
                 ${\tilde U}^{^1}$  \\
                 ${\tilde U}^{^2}$  \\
                 ${\tilde U}^{^3}$  \\
                 ${\tilde U}^{^4}$  \\
                 \end{tabular} \right).
\end{eqnarray}
In this case, the matrix $a_{_{a}}^{^{^{~b}}}(g_{_U})$ is the same form as in \eqref{d.4}.
Calculating the matrices $M_{_{ab}}$, ${N^{^a}}_{_{b}}$ and ${\Pi}^{^{ab}}(g_{_U})$ for this decomposition we then get
\begin{eqnarray}
NM^{^{-1}} = \left( \begin{array}{cccc}
e^{-t}  & -1  & 0 & 0\\
1  & e^{-t}  & 0 & 0\\
0 & 0  & 0 & e^{t} -2\\
0  & 0  & -(e^{t} +2)& 0\\
\end{array} \right),~~~{\Pi}(g_{_U}) = \left( \begin{array}{cccc}
0  & 0  & 0 & 0\\
0  &0  & -\chi & 0\\
0 & \chi  & -4y & 0\\
0  & 0  & 0& 0\\
\end{array} \right),\label{d.16}
\end{eqnarray}
leading to a background
\begin{eqnarray}
{E}_{_{ab}}(g_{_U}) &=& \left( \begin{array}{cccc}
\frac{e^{^{-t }}}{\Delta(t)} & \frac{1}{\Delta(t)}  & 0 & -\frac{\chi}{(e^t +2)\Delta(t)}\\\vspace{1mm}
-\frac{1}{\Delta(t)}  & \frac{e^{{-t }}}{\Delta(t)}  & 0 & -\frac{\chi e^{{-t }}}{(e^t +2)\Delta(t)}\\\vspace{1mm}
0 & 0  & 0 & \frac{1}{(e^t +2)}\\\vspace{1mm}
-\frac{\chi}{(e^t -2) \Delta(t)}  & \frac{\chi e^{{-t }}}{(e^t-2) \Delta(t)}  & -\frac{1}{(e^t -2)} & \frac{4 y}{(e^{{2t}} -4)}\\\vspace{1mm}
\end{array} \right),\nonumber\\
\Phi_{_U} &=&\phi^{^{(0)}} -2t + \frac{1}{2} \log\big(\frac{e^{^{2t}} -4 }{\Delta(t)}\big).\label{d.17}
\end{eqnarray}
In order to evaluate the total dilaton contribution we have to use $\phi^{^{(0)}}= t+\varphi_{_{0}}$
which gives the final result
\begin{eqnarray}\label{d.18}
\Phi_{_U} = \varphi_{_{0}} -t + \frac{1}{2} \log\big(\frac{e^{2t} -4 }{\Delta(t)}\big).
\end{eqnarray}
To go to the coordinate basis we have to use the left-invariant one-forms \eqref{c.11} to write the
background as
\begin{eqnarray}
ds^2 &=& -dt^2 +\frac{e^{^{-t }}}{\Delta(t)} \Big[dy^2 +dx^2 +\frac{2 \chi e^{2t}}{e^{^{2t}}-4} ~dy d \chi +
\frac{\chi (e^{2t}-8)}{e^{^{2t}}-4} ~dx d \chi\Big] -\frac{2e^{t}}{e^{^{2t}}-4} d \psi d \chi,\nonumber\\
B&=&\frac{1}{\Delta(t)}\Big[dy \wedge dx +\frac{\chi (e^{2t}-8)}{2(e^{^{2t}}-4)} dy \wedge d \chi +
\frac{\chi}{e^{^{2t}}-4} dx \wedge d \chi\Big]\nonumber\\
&&~~~~~~~~~~~~~~~~~~~~~~~~~~~~~~~~~~~~~~~~~~~~~~+\frac{2}{e^{^{2t}}-4} \big(d \psi \wedge d \chi  -2y d \chi \wedge d \chi\big),\nonumber\\
{\Phi} (t) &=& \varphi_{_{0}} -t + \frac{1}{2} \log\big(\frac{e^{^{2t}} -4 }{\Delta(t)}\big).\label{d.19}
\end{eqnarray}
By taking into consideration the scalar curvature of the metric which is
\begin{eqnarray}
R=-50 \frac{(e^{2t} +4e^{-2t}-2)}{(e^{2t} -4)^2 \Delta^{^2}(t)},\label{d.20}
\end{eqnarray}
one verifies the one-loop conformal invariance conditions of the background \eqref{d.19} with vanishing cosmological constant.
We thus studied a concrete example of the super PL T-plurality. Starting with the double $\big(({\C}^3 +{\A}) , {\cal I}_{_{(2|2)}}\big)$
we found the conformal duality chain of $2+1$-dimensional cosmological string backgrounds coupled with two fermionic fields in the form
of equations \eqref{d.6}, \eqref{d.13} and \eqref{d.19}.


\subsection{Conformal duality chain starting from $\big(gl(1|1) , {\cal I}_{_{(2|2)}}\big)$}

\smallskip
$\bullet$~$\big(gl(1|1) , {\cal I}_{_{(2|2)}}\big)$:\\\\
\smallskip
We take $\big(gl(1|1) , {\cal I}_{_{(2|2)}}\big)$ as the canonical decomposition. The corresponding Lie superalgebra is given by \cite{ER6}
\begin{eqnarray}
[T_{_1} , T_{_3}] &=&T_{_3},~~~~~~~~~~~[T_{_1} , T_{_4}]=-T_{_4},~~~~~~~~~~\{T_{_3} , T_{_4}\}=T_{_2},\nonumber\\
{[T_{_1} , {\tilde T}^{^3}]}&=&-{\tilde T}^{^3},~~~~~~~~[T_{_1} , {\tilde T}^{^4}]= {\tilde T}^{^4},~~~~~~~~~~~~[{\tilde T}^{^2} , T_{_3}]= {\tilde T}^{^4},\nonumber\\
{[{\tilde T}^{^2} , T_{_4}]}&=&{\tilde T}^{^3},~~~~~~~~~~
\{T_{_3} , {\tilde T}^{^3}\}=-{\tilde T}^{^1},~~~~~~~~~\{T_{_4} , {\tilde T}^{^4}\}={\tilde T}^{^1}. \label{d.21}
\end{eqnarray}
Choosing the parametrization of a general group element of $GL(1|1)$ as in \eqref{c.9} and employing
\eqref{b.4} we obtain
\begin{eqnarray}\label{d.22}
a_{_{a}}^{^{^{~b}}}(g) =\left( \begin{array}{cccc}
1  & -\psi \chi e^y  & -\psi &  \chi e^y\\
0  & 1  & 0 & 0\\
0 & -\chi  & -e^{-y} & 0\\
0  & -\psi e^y   & 0 & -e^y\\
\end{array} \right),~~~~~~~~~~~~b^{^{ab}}(g)=0,
\end{eqnarray}
then, it is simply followed that  $\Pi (g) =0$.
Also, in the canonical decomposition we have ${N^{^a}}_{_{b}} = {\delta^{^a}}_{_{b}}$ and $M_{_{ab}} = {E_{_{0}}}_{ab}$
and thus $E =E_{_0}$ as this was expected.
The dilaton field is found by the use of equation \eqref{b.18} to be
$\Phi = \phi^{^{(0)}}$. Here we want the total dilaton to be constant, so we
choose $\phi^{^{(0)}} = \varphi_{_{0}}$.
Our goal is now to obtain the background \eqref{c.21} from
a $\sigma$-model constructing on the double $\big(gl(1|1) , {\cal I}_{_{(2|2)}}\big)$. To do so,
we have to choose  the $E_{_0}$ as follows:
\begin{eqnarray}\label{d.23}
{E_{_0}}_{ab}(t) =\left( \begin{array}{cccc}
-\frac{1}{2} t^2 & b_{_0}  & 0 & 0\\
b_{_0}  & 0  & 0 & 0\\
0 & 0  & 0 &b_{_0}\\
0  & 0  & -b_{_0} & 0\\
\end{array} \right),
\end{eqnarray}
where $b_{_0}$ is a constant that differs from $\{0, 1, -1\}$.
Inserting \eqref{d.23} and the corresponding left-invariant one-forms given by equation \eqref{c.17} into action \eqref{c.6.1}
one gets
\begin{eqnarray}
ds^2 &=& -dt^2 - \frac{1}{2}  t^2 dy^2 + 2 b_{_0} dy dx -2 b_{_0} e^{y} d \psi d \chi,\nonumber\\
B&=&0,\nonumber\\
{\Phi} (t) &=& {\varphi}_{_0},\label{d.24}
\end{eqnarray}
which is nothing but the background \eqref{c.21}.
\\
\\
$\bullet$~$ \big(gl(1|1) , {\C}_{_{p=-1}}^{^{2}} \oplus {\A}_{_{1,1}}.ii\big)$:~~~~~~~~~\\\\
The Lie superalgebra of the double $ \big(gl(1|1) , {\C}_{_{p=-1}}^{^{2}} \oplus {\A}_{_{1,1}}.ii\big)$
obeys the following set of non-trivial (anti)commutation relations \cite{ER6}:
\begin{eqnarray}
[U_{_1} , U_{_3}] &=&U_{_3},~~~~~~~~~~~~~[U_{_1} , U_{_4}]=-U_{_4},~~~~~~~~~~~~~\{U_{_3} , U_{_4}\}=U_{_2},\nonumber\\
{[{\tilde U}^{^2} , {\tilde U}^{^3}]}&=&{\tilde U}^{^3},~~~~~~~~~~~~~[{\tilde U}^{^2} , {\tilde U}^{^4}]=- {\tilde U}^{^4},
~~~~~~~~~~~~[U_{_1} , {\tilde U}^{^3}]= -{\tilde U}^{^3},\nonumber\\
{[U_{_1}  , {\tilde U}^{^4}]}&=&{\tilde U}^{^4},~~~~~~~~~~~~~[{\tilde U}^{^2} , U_{_4}]= U_{_4}+{\tilde U}^{^3},~~~~~~~~
[U_{_3} , {\tilde U}^{^2}]= U_{_3}-{\tilde U}^{^4},\nonumber\\
{\{U_{_3}  , {\tilde U}^{^3}\}}&=& U_{_2} - {\tilde U}^{^1},~~~~~~\{U_{_4} , {\tilde U}^{^4}\}= -(U_{_2} - {\tilde U}^{^1}). \label{d.25}
\end{eqnarray}
The decomposition matrix between the doubles $\big(gl(1|1) , {\cal I}_{_{(2|2)}}\big)$
and $\big(gl(1|1) , {\C}_{_{p=-1}}^{^{2}} \oplus {\A}_{_{1,1}}.ii\big)$ is given by \cite{ER6}:
\begin{eqnarray}\label{d.26}
\left( \begin{tabular}{c}
                 $T_{_1}$  \\
                 $T_{_2}$  \\
                 $T_{_3}$  \\
                 $T_{_4}$  \\
                 ${\tilde T}^{^1}$  \\
                 ${\tilde T}^{^2}$  \\
                 ${\tilde T}^{^3}$  \\
                 ${\tilde T}^{^4}$  \\
                 \end{tabular} \right)=\left( \begin{tabular}{cccccccc}
                 1 & 0  & 0 & 0 & 0 & 0 & 0 & 0\\
                 0 & 1  & 0 & 0 & 0 & 0 & 0 & 0\\
                 0 & 0  & -1 & 0 & 0 & 0 & 0 & 0\\
                 0 & 0  & 0 & -1 & 0 & 0 & 0 & 0\\
                 0 & -1  & 0 & 0 & 1 & 0 & 0 & 0\\
                 1 & 0  & 0 & 0 & 0 & 1 & 0 & 0\\
                 0 & 0  & 0 & 0 & 0 & 0 & -1 & 0\\
                 0 & 0  & 0 & 0 & 0 & 0 & 0 & -1\\
                 \end{tabular} \right) \left( \begin{tabular}{c}
                 $U_{_1}$  \\
                 $U_{_2}$  \\
                 $U_{_3}$  \\
                 $U_{_4}$  \\
                 ${\tilde U}^{^1}$  \\
                 ${\tilde U}^{^2}$  \\
                 ${\tilde U}^{^3}$  \\
                 ${\tilde U}^{^4}$  \\
                 \end{tabular} \right).
\end{eqnarray}
Employing \eqref{b.4}, \eqref{d.23} and \eqref{d.25} the matrices \eqref{b.14}, \eqref{b.15} and \eqref{b.16} are obtained for this decomposition, giving
\begin{eqnarray}\label{d.27}
M_{_{ab}} &=&\left( \begin{array}{cccc}
-\frac{1}{2} t^2   & b_{_0}  & 0 & 0\\
b_{_0}  & 0  & 0 & 0\\
0 & 0  & 0 & -b_{_0}\\
0  & 0  & b_{_0} & 0\\
\end{array} \right),~~~~~{N^{^a}}_{_{b}}=\left( \begin{array}{cccc}
1-b_{_0}  & 0  & 0 & 0\\
-\frac{1}{2} t^2  & 1+b_{_0}  & 0 & 0\\
0 & 0  & -1 & 0\\
0  & 0  & 0 & -1\\
\end{array} \right).\nonumber\\
{\Pi}^{^{ab}} (g_{_U}) &=& \left( \begin{array}{cccc}
0  & 0  & 0 & 0\\
0  &0  & \psi e^y & -\chi\\
0 & -\psi e^y  & 0 & 0\\
0  & \chi  & 0& 0\\
\end{array} \right)
\end{eqnarray}
Using equations \eqref{b.13} and \eqref{b.17} it is then followed that
\begin{eqnarray}
{E}_{_{ab}}(g_{_U}) &=& \left( \begin{array}{cccc}
\frac{1}{b^2_{_0}-1}\big(\frac{1}{2} t^2 + 2 b^3_{_0} \psi \chi e^y\big) & \frac{b_{_0}}{b_{_0} +1}  & \frac{b^2_{_0} \chi}{b_{_0} +1}
&\frac{b^2_{_0} \psi e^y}{b_{_0} +1}\\\vspace{2mm}
-\frac{b_{_0}}{b_{_0}-1}  & 0  & 0 & 0\\\vspace{2mm}
\frac{b^2_{_0} \chi}{b_{_0} -1}  & 0  & 0 & b_{_0}\\\vspace{2mm}
\frac{b^2_{_0} \psi e^y}{b_{_0} -1} & 0  & -b_{_0} & 0\\
\end{array} \right),\nonumber\\
\Phi_{_U} &=&\phi^{^{(0)}} - \frac{1}{2} \log\big|b^2_{_0}-1\big|.\label{d.28}
\end{eqnarray}
Finally,  background including the supersymmetric metric,
the super antisymmetric tensor field and dilaton field is, in the coordinate basis, read off
\begin{eqnarray}
ds^2 &=& -dt^2 +\frac{1}{b^2_{_0}-1} \Big[\big(\frac{t^2}{2} +2 b^3_{_0} \psi \chi (e^y +1)\big) dy^2
-2b_{_0} ~dy dx  \nonumber\\
&&~~~~~~~~~~~~~~~~~~-2 b^3_{_0} \chi ~dy d \psi -2 b^3_{_0} \psi e^y (e^y +1) ~dy d \chi\Big] -2b_{_0} e^y~ d \psi d \chi,\nonumber\\
B&=&\frac{b^2_{_0}}{(b^2_{_0}-1)}\Big[dy \wedge dx +  \chi ~ dy \wedge d \psi + \psi e^y (e^y +1) ~ dy \wedge d \chi\Big],\nonumber\\
{\Phi} (t) &=& \varphi_{_{0}}  - \frac{1}{2} \log\big|b^2_{_0}-1\big|.\label{d.29}
\end{eqnarray}
Starting from the decomposition of semi-Abelian DSD $(gl(1|1) , {\cal I}_{_{(2|2)}})$
we could calculate the super PL T-plural of cosmology invariant with respect to the $GL(1|1)$  Lie supergroup.
In fact, we obtained another conformally invariant duality chain of $2+1$-dimensional
cosmological string backgrounds coupled with two fermionic fields in the form
of equations \eqref{d.24} and \eqref{d.29}.


\section{Conclusion}
We have generalized the formulation of PL T-plurality from Lie groups to Lie supergroups,
more strictly speaking, from DDs to DSDs.
Using the left-invariant one-forms of
the $(C^3 +A)$  and $GL(1|1)$ Lie supergroups we have obtained some new cosmological
string backgrounds including $(2+1|2)$-dimensional metric,  a dilaton field at most a function of $t$ only and
vanishing torsion
which are indeed conformally invariant up to the one-loop order.
The metrics of these backgrounds are flat in the sense that their scalar curvature is zero (more precisely, constant).
We have then shown that the resulting backgrounds are equivalent to
the ones of T-dual $\sigma$-models constructing on semi-Abelian DSDs
$ (({\C}^3 +{\A}) , {\cal I}_{_{(2|2)}})$ and $(gl(1|1) , {\cal I}_{_{(2|2)}})$.
Most importantly, starting from the above-mentioned decompositions of semi-Abelian DSDs
we have found the conformal duality/plurality chains of $2+1$-dimensional cosmological string backgrounds coupled with two fermionic fields.
We have furthermore checked that the backgrounds obtained by super PL T-plurality remain conformally invariant at one-loop level.
Our current goal was to get better understanding of super PL
T-plurality through investigation of examples presented in section 4 of the results.

As mentioned at the beginning of section 3, the Bianchi-type string cosmologies could be generalized to
$4 + 1$-dimensional cosmological models whose spatial
hypersurfaces are (simply) connected homogeneous Riemannian
manifolds \cite{Hervik}.
Our results can be also generalized to higher dimensions.
One may employ the formulation of super PL T-plurality in order to obtain the conformal duality chains of cosmological string backgrounds
in higher dimensions of the type $(m+1|2n)$. For instance, in dimension five one must use
the decompositions of DSDs generated by  Lie super bi-algebras of the type
$(3|2)$ \cite{Juszczak}. We intend to address this problem in the future.

In the introduction section, we  announced that the discussed super PL T-plurality
can yield insights to supergravity.
Our backgrounds have been constructed by the left-invariant one-forms (supervielbeins) on low-dimensional Lie supergroups of
the type $(2|2)$. One may use a Lie supergroup in higher dimensions of
the type $(10|32)$  as superspace to construct a general type II supergravity background of the Green-Schwarz superstring action \cite{L.Wulff1}.
Then, by using the formulation of super PL T-plurality, one can get the
duality/plurality of type II supergravity background.
Of course, for a generic type II Green-Schwarz superstring whose isometries contain a Lie supergroup G,
the transformation rules for the supergravity background fields under
non-abelian T-duality with respect to $G$  have already been worked out in \cite{L.Wulff2}.

However, we  don't know at the moment whether the resulting
backgrounds have other meaningful physical interpretation.
But, we hope that in future it will be possible to
find  super PL T-plural models even for physically interesting metrics.


\subsection*{Acknowledgements}

The author is especially grateful to A. Mehrvand for his careful reading of the manuscript.


\appendix

\section{Some properties of supermatrices and tensors on supervector
space}

In this appendix we give a few relevant details concerning properties of matrices and tensors on supervector space which
feature in the main text, appear as  supertranspose,  superdeterminant, supertrace, etc \cite{D}.

Let $\mathbb{G}$ be a supervector space with the bases ${\bf e}_{_M}$, and ${\bf e'}_{_N}$ be its dual bases.
The transformation between the bases can
be written as follows:
\begin{eqnarray}
{\bf e}_{_M}=(-1)^{^N}~ {K_{_M}}^{^N}~ {\bf e'}_{_N}.
\end{eqnarray}
One may consider the standard bases for the
supervector space $\mathbb{G}$ such that
in writing the bases as a column
matrix, one first presents the bosonic bases, then the fermionic
ones. In this way, the transformation matrix $K$ has the
following block diagonal representation \cite{D}
\begin{eqnarray}\label{A.2}
K=\left( \begin{tabular}{c|c}
                 A & C \\ \hline
                 D & B \\
                 \end{tabular} \right),
\end{eqnarray}
where $A,B$ and $C$ are real sub-matrices while $D$ is pure
imaginary sub-matrix.  Here we consider the matrices and
tensors having a form with all upper and lower indices written in
the right hand side.
The transformation properties of upper and lower
right indices to the left one are, for general tensors, given by
\begin{eqnarray}
^pT_{mn...}^{\;q}=T_{mn...}^{pq},\qquad
_nT^{pq}_{l...}=(-1)^n\;T_{nl...}^{pq}.
\end{eqnarray}
Let $K, L, P$ and $Q$ be the matrices whose elements indices have different positions. Then,
we define the {\it supertranspose } for these matrices in the following forms
\begin{eqnarray}
({{K^{st}})^{^M}}_{_N}&= &(-1)^{^{MN}}\;{K_{_N}}^{{^M}},~~~~~~~~~
{(L^{st})_{_M}}^{^N}=(-1)^{^{MN}}\;{L^{^N}}_{_M},\nonumber\\
(P^{st})_{_{MN}}&= & (-1)^{^{MN}}\;P_{_{NM}},~~~~~~~~~~
(Q^{st})^{^{MN}}=(-1)^{^{MN}}\;Q^{^{NM}}.
\end{eqnarray}
For the matrix  $K$  whose elements, $_{_M}K^{^N}$,  have the left index in the lower
position and the right index in the upper position, one defines  the {\it supertrace} as
\begin{eqnarray}
str K\;=\; (-1)^{^M} \;_{_M}K^{^M} = {K_{_M}}^{^M}.
\end{eqnarray}
When the matrix  $K$  is expressed in the block form  \eqref{A.2} the supertrace becomes
\begin{eqnarray}
str K\;=\; tr A- trB,
\end{eqnarray}
where ``tr'' denotes the ordinary trace. \\
If the sub-matrix $B$ in the block form  \eqref{A.2}  be a non-singular one, then the {\it superdeterminant}
for the matrix $K$ is defined by
\begin{eqnarray}
sdet\left( \begin{tabular}{c|c}
                 A & C \\ \hline
                 D & B \\
                 \end{tabular} \right)=det{(A-CB^{-1}D)}(det B)^{-1},
\end{eqnarray}
and if the sub-matrix $A$ be non-singular, then
\begin{eqnarray}
sdet\left( \begin{tabular}{c|c}
                 A & C \\ \hline
                 D & B \\
                 \end{tabular} \right)=\big(det{(B-DA^{-1}C)}\big)^{-1}\;(det A).
\end{eqnarray}
If both $A$ and $B$ are non-singular, then the {\it inverse}
matrix for \eqref{A.2} has the following form:
\begin{eqnarray}
{\footnotesize \left( \begin{tabular}{c|c}
                 A & C \\ \hline
                 D & B \\
                 \end{tabular} \right)^{-1}=\left( \begin{tabular}{c|c}
                 $(1_m-A^{-1}C
                  B^{-1}D)^{-1}A^{-1}$&
                  $-(1_m-A^{-1}CB^{-1}D)^{-1}A^{-1}CB^{-1}$  \\ \hline\vspace{1mm}

                 $-(1_n-B^{-1}DA^{-1}C)^{-1}B^{-1}DA^{-1}$  & $(1_n-B^{-1}DA^{-1}C)^{-1}B^{-1}$
                \end{tabular} \right),}
\end{eqnarray}
where  $m$ and $n$ are dimensions of
sub-matrices $A$ and $B$, respectively.

\smallskip
If $F(x)$ be a differentiable function on ${\mathbb{R}}_c^m \times {\mathbb{R}}_a^n$
where ${\mathbb{R}}_c^m$ are a subset of all real numbers ($c$-numbers) with dimension $m$ and ${\mathbb{R}}_a^n$ are a subset of all
odd Grassmann variables ($a$-numbers)  with dimension $n$, then relation between the left and right partial differentiations is given by
\begin{eqnarray}\label{A.10}
\frac{\overrightarrow{\partial}}{{\partial} x^{^M}} F \;=\; (-1)^{^{M(|F|+1)}}\; F \frac{\overleftarrow{\partial}}{{\partial} x^{^M}},
\end{eqnarray}
where $|F|$ indicates the grading $F$.\\
If $f$ be a scalar field, $\overrightarrow{\mathbf{V}}\;=\;V^{^M} \frac{\overrightarrow{\partial}}{{\partial} x^{^M}}$
a contravariant vector field and $\mathbf{\omega}\;=\; \omega_{_M} dx^{^M}$ a covariant vector field, then one finds {\it covariant derivative } in explicit components form as follows:
\begin{eqnarray}
f {\overleftarrow{\nabla}}_{_M} &=&  (-1)^{^{M|f|}}\; \overrightarrow{\nabla}_{_M} f = f \frac{\overleftarrow{\partial}}{{\partial} x^{^M}},\label{A.11}\\
V^{^M} {\overleftarrow{\nabla}}_{_N} &=&  (-1)^{^{N(|V|+M)}}\; \overrightarrow{\nabla}_{_N} ~V^{^M} = V^{^M} \frac{\overleftarrow{\partial}}{{\partial} x^{^N}}+(-1)^{^{P(M+1)}} ~V^{^P} {\Gamma}^{^M}_{_{\;~PN}},\label{A.12}\\
\omega_{_M} {\overleftarrow{\nabla}}_{_N} &=& (-1)^{^{N(|\omega|+M)}}\; \overrightarrow{\nabla}_{_N} ~\omega_{_M} = \omega_{_M} \frac{\overleftarrow{\partial}}{{\partial} x^{^N}}- \omega_{_P}~ {\Gamma}^{^P}_{_{\;~MN}},\label{A.13}
\end{eqnarray}
where ${\Gamma}^{^P}_{_{\;~MN}}$ are called the components of the connection $\bf{\nabla}$.\\
If the supersymmetric matrix ${_{_A}G}_{_B}$ (its inverse denotes to
${^{A}G}^{B}$, and ${G}^{^{AB}} = (-1)^{AB} {G}^{^{BA}}$) be the components
of metric tensor field on a Reimannian supermanifold, then, in a
coordinate basis, the components of the {\it connection} and {\it
Riemann tensor field} are given by
\begin{eqnarray}
~~~~~{\Gamma}^{^M}_{_{~NP}}&=& (-1)^{^Q}\;G^{^{MQ}} {\Gamma}_{_{QNP}}  \nonumber \\
&=& \frac{(-1)^{^Q}}{2} G^{^{MQ}}
\Big[G_{_{QN}}\frac{\overleftarrow{\partial}}{\partial
x^{^P}}+(-1)^{^{NP}}G_{_{QP}}\frac{\overleftarrow{\partial}}{\partial
x^{^N}}- (-1)^{^{Q(N+P)}}G_{_{NP}}\frac{\overleftarrow{\partial}}{\partial
x^{^Q}}\Big],~~~~~~~~~~~~~~~~\label{A.14}\\
R^{^I}_{_{~JKL}}&=& -{\Gamma}^{^I}_{_{~JK}}\frac{\overleftarrow{\partial}}{\partial
x^{^L}}+(-1)^{^{KL}}~{\Gamma}^{^I}_{_{~JL}}\frac{\overleftarrow{\partial}}{\partial
x^{^K}}+(-1)^{^{K(J+M)}}~{\Gamma}^{^I}_{_{~MK}} {\Gamma}^{^M}_{_{~JL}} \nonumber \\
&&~~~~~~~~~~~~~~~~~~~~~~~~~~~~~~~~~~~~~~~~~- (-1)^{^{L(J+K+M)}}~{\Gamma}^{^I}_{_{~ML}} {\Gamma}^{^M}_{_{~JK}}.\label{A.15}
\end{eqnarray}
In addition, for the {\it curvature tensor field}, the {\it Ricci
tensor} and the {\it curvature scalar field} we have
\begin{eqnarray}
R_{_{IJKL}}&=& G_{_{IM}} R^{^M}_{~_{JKL}},\label{A.16}\\
R_{_{IJ}}&=&(-1)^{^{K(I+1)}}\;R^{^K}_{_{\;~IKJ}}, \label{A.17}\\
R &=& {R_{_M}}^{^M} = {str}(R_{_{MN}} G^{^{NM}}).\label{A.18}
\end{eqnarray}
Accordingly, one may use the above definitions and formulas to rewrite the one-loop beta-function equations \eqref{c.3}-\eqref{c.5}
in the following form
\begin{eqnarray}
&&{R}_{_{MN}}-\frac{1}{4} (-1)^{^{D+BN}} H_{_{MBA}} G^{^{AD}} H_{_{DNS}} G^{^{SB}}~~~~~~~~\nonumber\\
&&~~~~~~~~~~~~~~~~~~~~~~~~~~~~~~~+2\Big[(-1)^{^{M+N}} \frac{\overrightarrow{\partial}}{\partial
x^{^{M}}} \big(\frac{\overrightarrow{\partial}\Phi}{\partial
x^{^{N}}}\big) -(-1)^{^{K}} \frac{\overrightarrow{\partial} \Phi}{\partial
x^{^{K}}} ~\Gamma^{^{K}}_{_{MN}}\Big] ~=~0,\label{a.5.1}\\
&&(-1)^{^{1+M+Q}} ~2\frac{\overrightarrow{\partial}\Phi}{\partial
x^{^{Q}}} ~G^{^{QM}} H_{_{MNP}} + (-1)^{^{M+L+LN+LP}}~G^{^{LM}}\Big[(-1)^{^{L(1+M+N+P)}}~\frac{\overrightarrow{\partial}}{\partial
x^{^{L}}} H_{_{MNP}}~~~~\nonumber\\
&&~~~~~~~~~-(-1)^{^{(P+N)(M+Q)}}~H_{_{QNP}} ~\Gamma^{^{Q}}_{_{ML}}-(-1)^{^{P(N+Q)}}~H_{_{MQP}} ~\Gamma^{^{Q}}_{_{NL}}-H_{_{MNQ}} ~\Gamma^{^{Q}}_{_{PL}}\Big] ~=~0,~~~~~~~~\label{a.6.1}\\
&&4 \Lambda-{R}+\frac{1}{12} (-1)^{^{M+BN}} H_{_{CBA}} G^{^{AM}} H_{_{MNP}} G^{^{PB}} G^{^{NC}} +4 (-1)^{^{M}}
(\frac{\overrightarrow{\partial} \Phi}{\partial x^{^{M}}}) G^{^{MN}} (\frac{\overrightarrow{\partial} \Phi}{\partial x^{^{N}}})\nonumber\\
&&~~~~~~~~~~~~~~~~~~~~~~~~~-4 G^{^{NM}} \frac{\overrightarrow{\partial}}{\partial
x^{^{M}}} \big(\frac{\overrightarrow{\partial}\Phi}{\partial
x^{^{N}}}\big)+4 (-1)^{^{K+M+N}}  G^{^{NM}} (\frac{\overrightarrow{\partial}\Phi}{\partial
x^{^{K}}}) \Gamma^{^{K}}_{_{MN}}~=~0,\label{a.7.1}
\end{eqnarray}
where ${R}_{_{MN}}$ and $R$ are defined according to equations \eqref{A.17} and \eqref{A.18}, respectively.
It's worth mentioning that all the lowering and raising of the indices  will be done with respect to the tensor fields $G_{_{MN}}$ and ${G}^{^{MN}}$.
As an example, for the components of tensor $T$ we have
\begin{eqnarray}
{{T_{_{A_{_1},\cdots, A_{_r}}}}^{^N}}_{B_{_1},\cdots, B_{_S}}\;=\;(-1)^{^{(M+N)(B_{_1}+\cdots+ B_{_S})}}~
T_{_{A_{_1},\cdots, A_{_r}, M, B_{_1},\cdots, B_{_S}}} {G}^{^{MN}}.
\end{eqnarray}


\end{document}